\documentclass[conference]{IEEEtran}
\usepackage{blindtext, graphicx}
\ifCLASSINFOpdf
\else
\fi

\usepackage[caption=false,font=footnotesize]{subfig}

\usepackage{fixltx2e}

\begin{document}

\title{Group Mobility: Detection, Tracking and Characterization}

\author{\IEEEauthorblockN{Ivan Oliveira Nunes}
\IEEEauthorblockA{Department of Computer Science\\
Federal University of Minas Gerais\\
Belo Horizonte, Minas Gerais\\
ivanolive@dcc.ufmg.br}
\and
\IEEEauthorblockN{Pedro O. S. Vaz de Melo}
\IEEEauthorblockA{Department of Computer Science\\
Federal University of Minas Gerais\\
Belo Horizonte, Minas Gerais\\
olmo@dcc.ufmg.br}
\and
\IEEEauthorblockN{Antonio A. F. Loureiro}
\IEEEauthorblockA{Department of Computer Science\\
Federal University of Minas Gerais\\
Belo Horizonte, Minas Gerais\\
loureiro@dcc.ufmg.br
}
}

\maketitle

\begin{abstract}

In the era of mobile computing, understanding human mobility patterns is crucial in order to better design protocols and applications. Many studies focus on different aspects of human mobility such as people's points of interests, routes, traffic, individual mobility patterns, among others. In this work, we propose to look at human mobility through a social perspective, i.e., analyze the impact of social groups in mobility patterns. We use the MIT Reality Mining proximity trace to detect, track and investigate group's evolution throughout time. Our results show that group meetings happen in a periodical fashion and present daily and weekly periodicity. We analyze how groups' dynamics change over day hours and find that group meetings lasting longer are those with less changes in members composition and with members having stronger social bonds with each other. Our findings can be used to propose meeting prediction algorithms, opportunistic routing and information diffusion protocols, taking advantage of those revealed properties.

\end{abstract}


\IEEEpeerreviewmaketitle

\section{Introduction}

Many practical problems can benefit from the knowledge of the underneath dynamics that govern human mobility. For example, it can be applied to better plan urban infrastructure, forecast traffic, map the spread of biological viruses, or better design Mobile Ad-hoc Network (MANET) protocols.

Specific network problems such as opportunistic routing and information diffusion in MANETs share a common interesting property: they are highly dependent on how humans interact with each other. In this context, we define a human group as set of people that, for some reason or goal, get together in space and time. It is clear that knowledge of regular group meetings can be explored to improve the current state of the art of opportunistic information diffusion, routing or to increase the current understanding of how diseases spread. However, it remains a challenge how to define, detect, keep track and analyze groups of humans and their dynamics.

In the literature, there are several proposals dedicated to understanding and modeling human mobility considering diverse aspects, but group mobility is currently an untrodden field. The human species is a pretty sociable and this sociability must be considered in order to better understand, model and predict movement. The growing ubiquity of geo-localization sensors and the availability of data collected by them create a new opportunity to tackle the problem of group mobility. With this in mind, the present work focuses on characterization of groups' dynamics through the analysis of proximity contact traces. Among the contribution of this work, we highlight:

\begin{itemize}
\item Definition of a methodology for telling apart random and social interactions in proximity traces using a time dependent social graph model;
\item Proposal of a systematic way for detecting and tracking human mobile groups;
\item Characterization of group mobility properties, including group evolution, periodicity and meeting durations;
\item A discussion and some early results on how knowledge of group mobility could be applied to design opportunistic network protocols.
\end{itemize}

This paper is organized as follows. Section \ref{related} reviews some of the related work and discusses the corresponding contributions to characterization and modeling of group mobility. Section \ref{metho} formalizes our methodological steps to detect track and characterize human groups' dynamics. Section \ref{char} describes the experiments' methodology and metrics, presenting results and the main groups' characteristics detected. Section \ref{case} discusses the application of group detection to information dissemination protocols. Finally, Section \ref{conclusion} brings the final remarks and future work.

\section{Related Work}\label{related}

\textit{Human mobility.} Some rules that govern human mobility have already been revealed. Gonzalez et {al.}~\cite{individual_mob} use a Call Detail Records (CDR) large scale data set to characterize individual mobility patterns among cellphone users in Europe. Their results show that the radial displacement in human trajectories, namely radius of gyration, follows a power-law with exponential cut. Their analysis also reveals that human trajectories are highly periodical. Some studies evaluate the impact of large scale events in urban mobility. Calabrese et {al.}~\cite{events} analyzed over 1 million cellphone records to unveil the impact of those kind of events in the city transit and mobility. They claim that the comprehension of the origin of people attending a given event is fundamental to propose policies for traffic congestion mitigation. Silva et {al.}~\cite{2014revealing} proposed a visualization technique, namely City Image, which captures typical transitions between PoIs in a city using publicly available data from Foursquare. ``City Images'' can be understood as insights to the mobility patterns of each specific city. There are several studies on characterization and prediction of traffic conditions using various types of data sources such as \cite{traffic1,traffic2,traffic3}. Some of the studies investigate how large scale events, like weather changes for example, may affect traffic conditions \cite{weather}. To the best of our knowledge, there is no such characterization related to group mobility, which is the focus of this work.

\textit{Community Detection.} To perform group characterization from mobility traces' analysis, we apply community detection methods. Since its introduction, community detection in complex networks have attracted a lot of attention. Algorithms for community detection can be classified according to two characteristics: overlapping versus non-overlapping, and static versus dynamic graphs. Among many proposed algorithms, the studies in \cite{palla2005} and \cite{copra} have remarked themselves as the most popular and effective algorithms for community detection in static graphs. Studies such as \cite{c2} aim to propose adaptations and new algorithms that are suited for dynamics graphs, considering computational efficiency issues. In our study, we use these developed methodologies, specifically \textit{Clique Percolation} \cite{palla2005}, to detect and characterize social groups dynamics looking at proximity traces.

\textit{Community Characterization.} There are also studies that characterize community evolution in other kinds of complex networks, but not in mobility networks. For instance, Palla et {al.} \cite{palla2007} analyzed the evolution of communities in scientific collaboration networks and in phone call networks. Hui et {al.} \cite{bubblerap} used community structure in mobility networks to design a very successful message forwarding protocol for Disruption Tolerant Networks (DTN), namely Bubble Rap. Our work is fundamentally different from \cite{bubblerap} because it aims to detect groups of people who are in fact together, in space and time, socially interacting. In \cite{bubblerap}, the authors build a single static graph, of the whole trace time, and detect static communities in this single graph. The use of a single aggregated graph prevents accounting for changes of social behavior in time, i.e., social dynamics. For example, students who meet regularly because they attend the same class in a given semester may not be attending a class together in the next semester. The methodology we propose is capable of accounting for these changes in behavior through the detection and tracking of social groups who are together in space and time, instead of social communities in aggregated graphs. In this paper, we aim to characterize the evolution of social groups by looking at properties such as group sizes, group meeting durations, periodicity in group meetings and dynamics of groups' evolution. Finally, we discuss group mobility application, providing a case study and some early results for opportunistic mobile networks.

\section{Social Groups Identification and Tracking}\label{metho}

In this section, we go over the methodological steps we performed to be able to identify and track social groups from proximity traces.

\subsection{Modeling the Evolution of Proximity Traces With Graphs}\label{model}

To analyze group dynamics we propose a social graph model. Firstly, we slice the proximity data set in time windows $tw$ (we discuss the ideal time size for $tw$ in Section \ref{data_cha}). Contacts within the same slice $tw$ will be aggregated in a contact graph $Gc(V,E[tw = i])$, in which $V$ is the set of vertices representing entities in the data set (i.e., people) and $E$ is the set of edges that represent proximity contacts between a pair of entities in $V$. Thus, in our model, the trace processing will result in a set of subsequent, undirected, edge-weighted graphs: $S = \{Gc(V,E[tw=0]),Gc(V,E[tw=1]), ..., Gc(V,E[tw =n])\}$. The weight of an edge $(v,w)$ $\in$ $E$ is given by (i) the number of contacts registered between $v$ and $w$ during the time slice $tw$; or (ii) the sum of $(v,w)$ contacts durations, inside $tw$ time slice, if contacts' durations are available in the trace. In this paper, we used a trace in which contact durations were not available, so we computed edge weights according to the first option.

\subsection{Telling Apart Social and Random Contacts to Create Social Graphs}\label{random}

\begin{figure}[!t]
\centering
\includegraphics[width=\columnwidth]{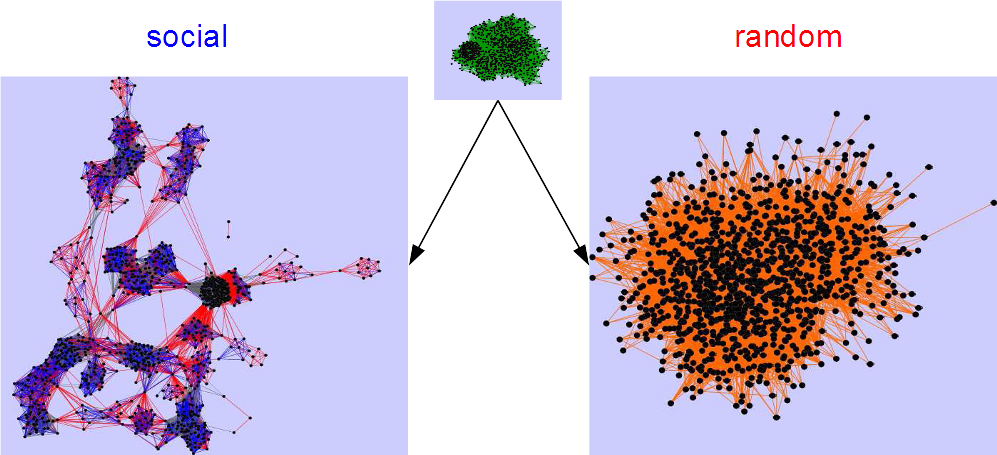}
\caption{RECAST application to tell apart random and social relationships between peers in a contact trace}
\label{recast}
\end{figure}

To help separating random and social contacts, we apply RECAST algorithm \cite{recast} to the contact trace. RECAST algorithm separates social from random relationships between peers (Fig.\ \ref{recast}). It works comparing the edge persistence and topological overlap of randomly generated graphs to the actual contact graph obtained from the contact trace. As output, RECAST reveals which pairs of nodes meet each other in a social fashion and each pairs do not share social properties. We then remove from the trace, contacts between pairs classified by RECAST as random and proceed with the analysis in the trace containing only contacts between pairs that share social bonds.

Considering a pre-defined $tw$ size, for instance 30 minutes, a single contact within the whole time slice does not necessarily mean that the entities are socially interacting. Single contacts might be random encounters, even if nodes share a social bond. It might be caused by intersections between individual trajectories and should not necessarily mean a social interaction. We are interested in defining a threshold $w_{th}$ for the minimum edge weight, which should be enough to consider that the entities are in fact together inside the time slice $tw$. It is clear that to properly define both, the size of slice $tw$ and the edge weight threshold $w_{th}$, we need to analyze data set properties (e.g. sampling rate). In Section~\ref{data_cha}, we perform a characterization of the studied data set to be able to properly define $tw$ and $w_{th}$. The following methodology can be applied to other data sets as well.

\subsection{Data Set Characterization}\label{data_cha}

To be able to analyze the social groups' dynamic properties, we need to understand the data set, avoiding potential biases (e.g., sampling bias, and inconsistencies) due to the data acquisition process. We perform this evaluation with the goal of defining $tw$ size (the proper time slice to divide the data set) and $w_{th}$ (the threshold for the number of contacts, or contact duration that tell apart social and random contacts in the trace previously filtered by RECAST).

In the present study, we used the MIT Reality Mining proximity trace \cite{mit}, which is a contact trace containing 80 users who reside in two different university buildings. Users were monitored for more than two years and contacts were registered when two users were less than 10 meters apart. A contact entry in the trace is composed of the IDs of the pair of nodes and the date and time when the contact happened. It is worth mentioning that geo-location traces (such as GPS traces) can be converted to proximity traces by defining a minimum distance, which can be considered a contact between two entities. For this reason, this methodology can also be applied to those kinds of traces.

\begin{figure*}[!t!b]
\centering
\subfloat[]{\includegraphics[width=2.0in]{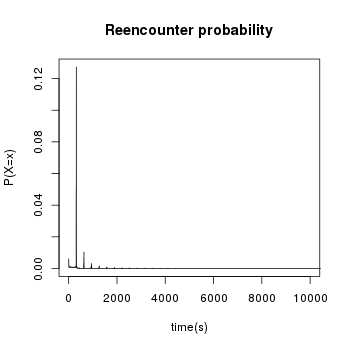}
\label{fig_first_case}}
\hfil
\subfloat[]{\includegraphics[width=2.0in]{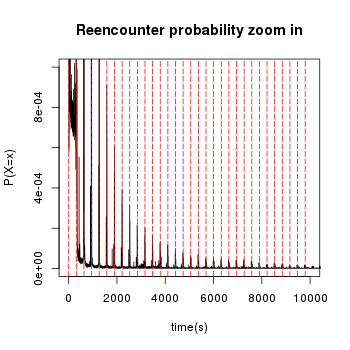}
\label{fig_second_case}}
\hfil
\subfloat[]{\includegraphics[width=2.0in]{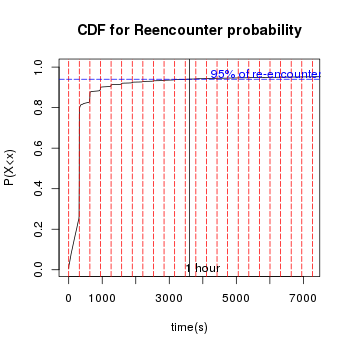}
\label{fig_third_case}}
\caption{Probability function of the time $x$ (in seconds) until the next meeting. Red dashed lines show a fixed inter-measurement time of 318 seconds in which re-encounter peaks happen. This means that most of the trace proximity records were acquired in fixed periods of 318 seconds. The probability of a pair of nodes meeting again has approximately exponential distribution and 95\% of the re-encounters happen in less than one hour}
\label{fig_reenc}
\end{figure*}

Firstly, we analyze the time between pair re-encounters, i.e., once a pair has met, what is the distribution of the time until the next meeting. Fig.\ \ref{fig_reenc} shows that the re-encounter behavior is very periodical, with peaks around every five minutes (red dashed lines). This behavior indicates that the deployed system for data acquisition acts every five minutes most of the times, but for some reason it can also actuate in shorter periods. Looking at the CDF of the re-encounter probability, we see that approximately 95\% of the re-encounters can be captured with a $tw$ of one hour. For this reason, we set the duration of the time window $tw$ to one hour.

\begin{figure}[!t]
\centering
\includegraphics[width=\columnwidth]{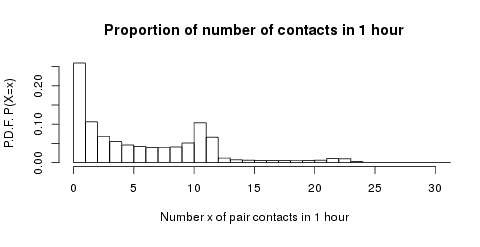}
\caption{Probability distribution function of the number of pair contacts per hour.}
\label{hour_contacts}
\end{figure}

Next, we analyze the fraction of pairs contacts to define $w_{th}$. Fig,\ \ref{hour_contacts} shows that 27\% of pairs that meet in a given hour only meet once. We assume these one-time meetings as coincidence meetings. For meeting frequencies from 2 to 12, the graphic shows values between 5\% and 10\%. For frequencies higher than 12, the probability becomes very low, which is consistent to the assumption of data acquisition mostly in periods of five minutes, but rarely less than five. From 2 to 12 encounters per hour, we have similar values in the PDF when compared to $P(X = 1)$. For this reason, $w_{th} = 2$ for the MIT Reality Mining data set.

To summarize, through the data set characterization, we were able to define that two or more contacts within an hour are enough to be considered social interaction (after RECAST filtering).

\begin{figure*}[!t]
\centering
\subfloat[7AM]{\includegraphics[width=2.0in]{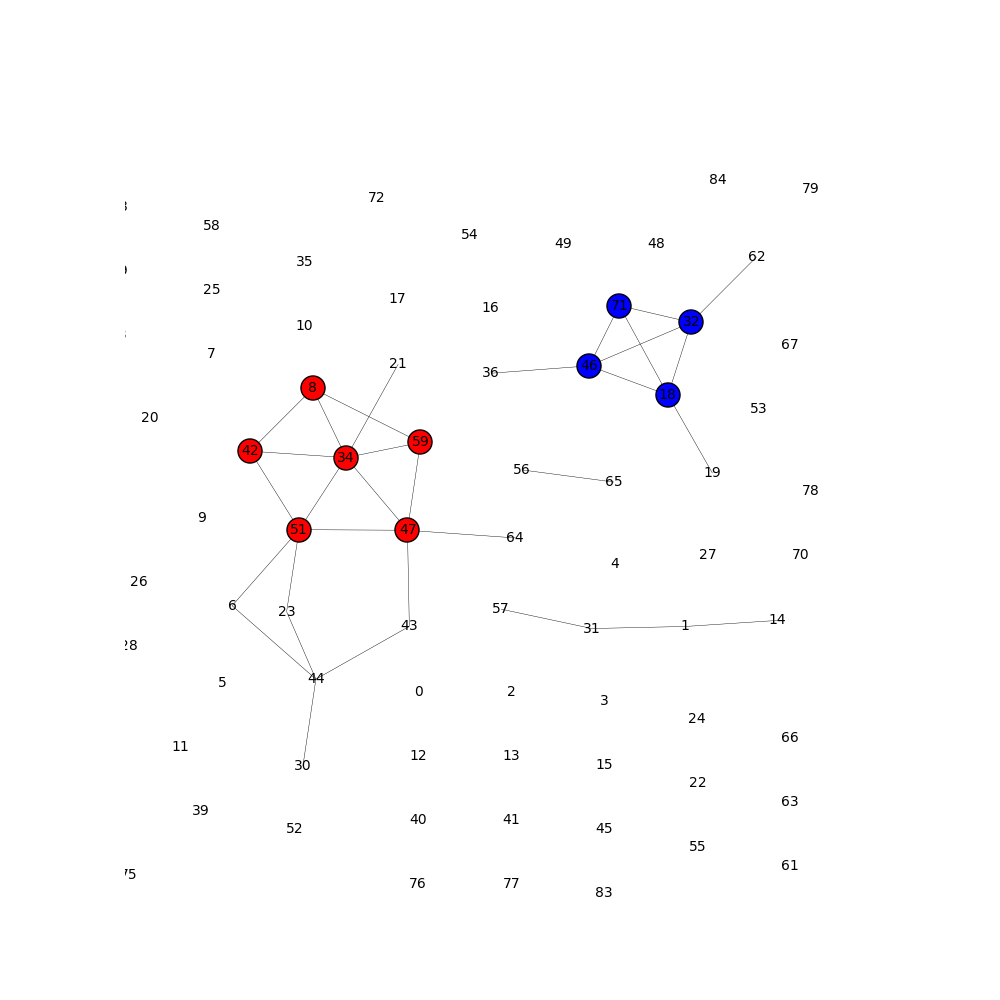}
\label{g1}}
\hfil
\subfloat[8AM]{\includegraphics[width=2.0in]{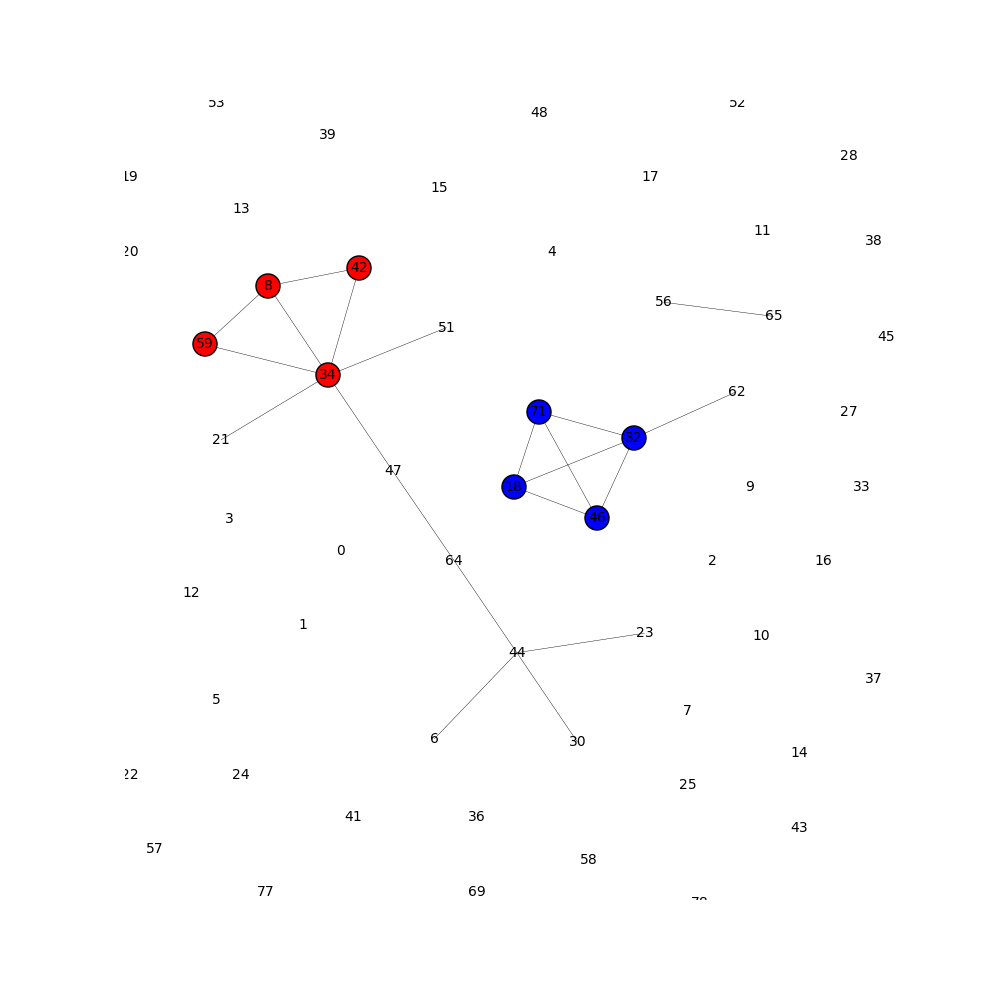}
\label{g2}}
\hfil
\subfloat[9AM]{\includegraphics[width=2.0in]{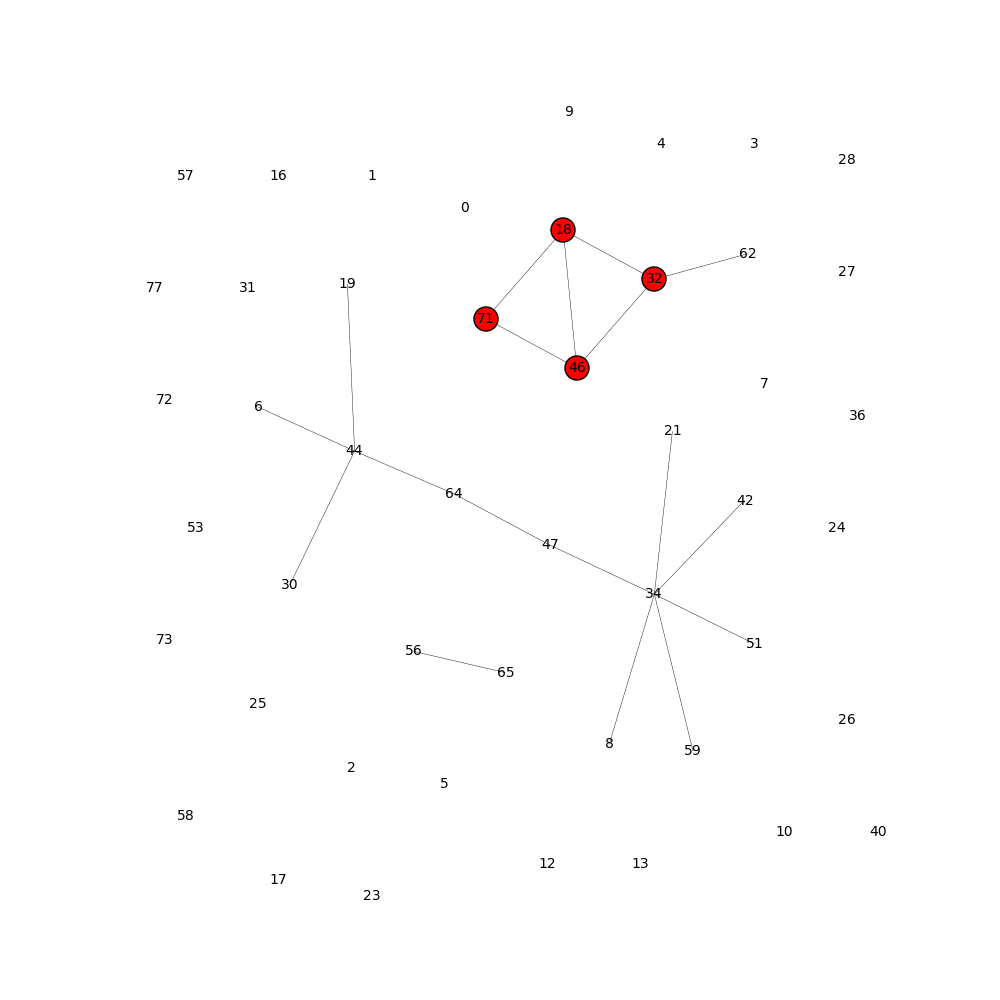}
\label{g3}}
\caption{Group detection with CPM, in the MIT proximity trace, with $tw = 1h$ in three consecutive time windows, at February 5th of 2009. Only edges with $w_{th} \geq 2$ are represented}
\label{mit_groups}
\end{figure*}

\subsection{Group Detection}\label{cpm}

After defining values for $tw$ and $w_{th}$, we define a social a group as follows:\\
\begin{itemize}
\item{\textbf{\textit{Definition of social group:}}}\textit{ A group is a community detected in $Gc(V,E[tw=i])$ , i.e., the graph generated from the $i^{th}$ time slice of the trace $S$, after eliminating contacts between pairs with no social bond and edges with weight bellow the threshold $w_{th}$.}
\end{itemize}

So far, we have established a model to represent social interactions that consists of graphs generated from peer contacts in traces' time slices. Following the above group definition, we must be able to detect communities (social groups represented by more densely interconnected parts within a graph of social links) in such graphs in order to track social groups. There are several community detection algorithms, such as \cite{c1,c2,copra}. From the existing algorithms, in this work we use the Clique Percolation Method (CPM) \cite{palla2005}. The main reasons for using CPM are that their community members can be reached through well connected subsets of nodes and that the communities may overlap (share nodes with each other). This latter property is essential, as most social graphs are characterized by overlapping and nested communities \cite{palla2007}. For each time-slice graph $Gc(V,E[tw=i])$ we compute CPM.

In CPM, a community is defined as a union of all k-cliques (complete sub-graphs of size $k$) that can be reached from each other through a series of adjacent k-cliques (where adjacency means sharing $k-1$ nodes). The CPM parameter $k$ limits the minimum size of detected communities. CPM has remarked itself as one of the most effective methods once fed with correct parameters \cite{parameters}. To the purpose of detecting social groups, we set $k=3$, thus, we consider groups of three or more people. Fig.\ \ref{mit_groups} shows groups detected with CPM in the MIT proximity trace \cite{mit} at February 5th of 2009, at 7, 8 and 9am, subsequent time slices of one hour size.

\subsection{Group Tracking}\label{track}

Once groups are detected in different time slices, there must be a way of tracking them, i.e., a criterion for considering that two groups in consecutive time slices are in fact the same group. With that goal we introduce the Group Correlation Coefficient $\rho(G1,G2)$:
\begin{equation}\label{stability}
    \rho(G1,G2) = \frac{|V(G1) \cap V(G2)|}{|V(G1) \cup V(G2)|}
\end{equation}
where $|V(G1) \cap V(G2)|$ is the number of common nodes in groups $G1$ and $G2$ and $|V(G1) \cup V(G2)|$ is the total number of  different nodes that compose both groups. The coefficient $\rho$ assumes values from 0 to 1, where 0 means no correlation, i.e., no node that belongs to both groups and 1 means that $G1$ and $G2$ have the exact same node composition. Group correlation coefficient is a measurement of the stability in groups' composition.

We consider a group to be the same in two consecutive time slices if $\rho(G(tw=i),G(tw=i+1)) > 0.5$, i.e., if at least 50\% of the group members remains the same. A $\rho$ value greater than $0.5$, is the condition to map each group in a single group in two different time slices. The value $\rho < 0.5$, would allow a single group to be mapped to two different groups with less than half of the original node composition in the next time window, adding complexity to the group tracking. At the same time a $\rho$ threshold of $0.5$ allows high volatility in group composition, making it possible to better analyze groups' evolution.

\section{Characterization of Group Dynamics}\label{char}

\begin{figure*}[!t]
\centering
\subfloat[Distribution of group sizes over day hours (0 to 23h)]{\includegraphics[width=2.0in]{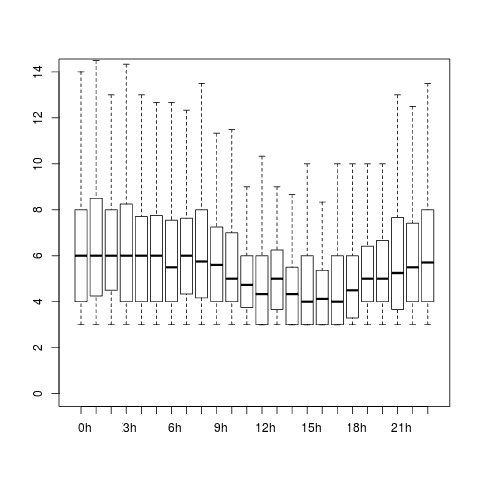}
\label{gd1}}
\hfil
\subfloat[Considered events in social group evolution]{\includegraphics[width=2.0in]{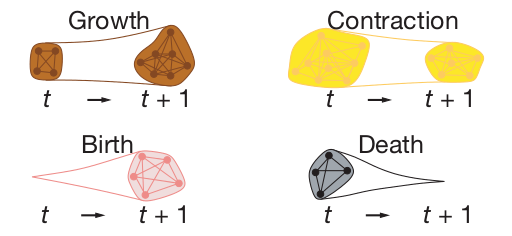}
\label{gd2}}
\hfil
\subfloat[Dynamics of group evolution throughout day periods]{\includegraphics[width=2.0in]{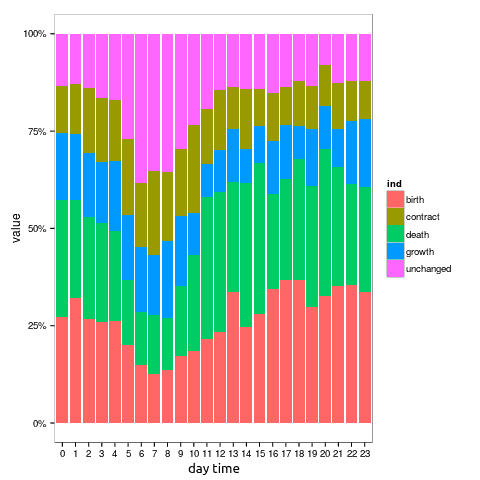}
\label{gd3}}
\hfil
\caption{Analysis of groups' evolution, i.e., groups' sizes and groups' transformations, over different days times.}\label{evoluttion}
\end{figure*}

\begin{figure}[!t]
\centering
\includegraphics[width=\columnwidth]{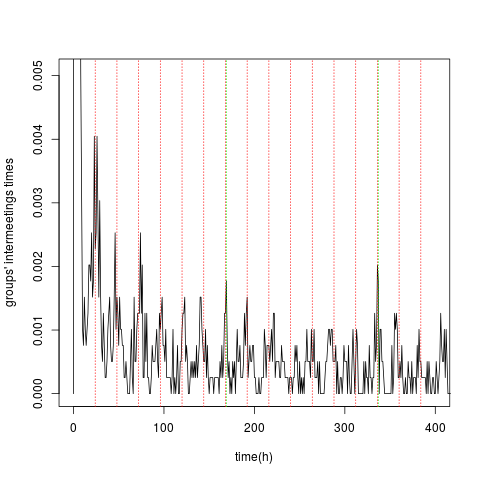}
\caption{Probability of a group re-meeting in $t$ hours after meeting for the first time at $t=0$. Red lines represent 24-hour periods and green lines represent 7-day periods}
\label{period}
\end{figure}

In this section we use the results of group detection and tracking methodology proposed in Section~\ref{metho} to characterize groups' dynamic evolution. Specifically, we analyze the following characteristics:
\begin{itemize}
\item Groups' sizes throughout day hours.
\item Groups' evolution considering possibilities of no change, growth, contraction, birth and death.
\item Groups' meetings inter-contact times and periodicity.
\item Groups' meetings durations and its correlation with group stability and with group's social bonds strength.
\end{itemize}

\subsection{Metrics}\label{metrics}

One of the interests of the experiments to be presented in this section, is to reveal what factors impact the duration of a group meeting. Specifically, we plan to investigate the impact of the stability of group members and the impact of the strength of social bonds shared by group members.

To measure the stability of group members we use the previously defined Group Correlation Coefficient $\rho$. Since we consider a group to remain the same if $\rho > 0.5$, the value for $\rho$ throughout the duration of a group may vary from $0.5$ to $1.0$.

For measuring the strength of groups' social bonds we define Groups Self-Containment Factor (GSCF) as:
\begin{equation}\label{gscf}
    GSCF(G) = \frac{\sum{w_{in}(G)}}{\sum{w_{in}(G)}+\sum{w_{out}(G)}}
\end{equation}
where $\sum{w_{in}(G)}$ is the sum of edge-weights between members of Group $G$, and $\sum{w_{out}(G)}$ is the sum of edge-weights between members of group $G$ and outsiders (non-members). A $GSCF(G)$ value close to one means that G members share strong social bonds while a value close to zero means a group with weak social links.

\subsection{Results}

After applying the methodology for group detection and tracking, proposed in Section \ref{metho}, we here analyze characteristics from detected groups. Figure \ref{gd1} shows box-plots for groups sizes are presented for each day hour (outliers omitted for better presentation). It shows that in night hours (10pm to 7am), groups' sizes are similar and distributed around the average size of six. During day hours (8am to 9pm), groups' sizes are more heterogeneous and have lower sizes. This behavior indicates that, in hours with higher mobility, groups tend to be less stable and have reduced number of members.

According to the considered possibilities of evolution in groups, displayed in Fig.\ \ref{gd2}, Fig.\ \ref{gd3} presents the average group dynamics throughout hours of the day. In late hours and early morning, the number of births and deaths decreases and the number of unchanged groups, i.e., groups that maintain the same number of nodes in consecutive hours, increases. This behavior is explained by the fact that late hours roommates and dorm neighbors probably form groups. Although few group members may leave or join, it is more probable that groups remain until the next working hours, when students leave dorms.

Fig.\ \ref{period} presents a very important result, which measures the frequency of group re-encounters, i.e., given the fact that a group first met at time $t=0$, how group re-meetings are distributed along the next hours ($t$ hours after the first meeting). The result reveals that the mass of probability is concentrated around peaks in periods of 24 hours (represented by red dashed lines). This means that group meetings are highly periodical in a daily fashion. One may also notice higher peaks marked with green dashed lines. Green dashed lines represent periods of 7 days, meaning that groups meetings also present weekly periodicity. This periodicity makes sense since people have schedules and routines. This result motivated our discussion in Section \ref{case}, which tries to answer the question: Is it possible to use past group meetings to predict future ones?

\begin{figure}[!t]
\centering
\includegraphics[width=\columnwidth]{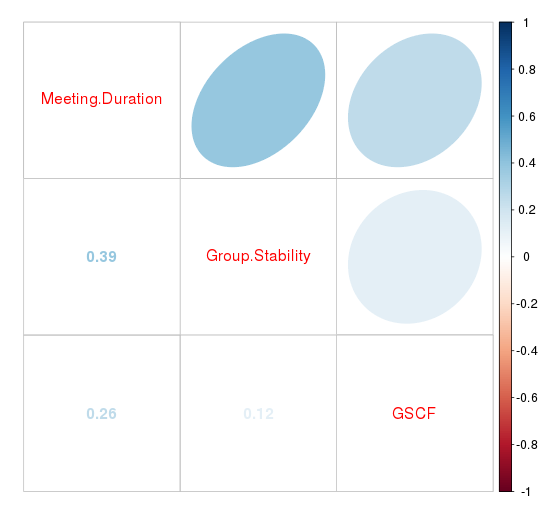}
\caption{Pearson's correlation between group meeting durations, $GSCF$ and members stability $\rho$.}
\label{correlations}
\end{figure}

To measure the impact of the strength of social bonds ($GSCF$ - Eq.\ 2, Section \ref{metrics}) and groups' stability ($\rho$, Eq.\ 1, Section \ref{track}) to the group meetings durations we compute Pearson's Correlation. Fig.\ \ref{correlations} shows that group meetings durations present moderate positive correlation with both, $GSCF$ and $\rho$, meaning that a condition for long group life times is that group composition remain mostly unchanged and the existence of strong social bonds between group members. The presented correlations were obtained with a p-value of $2.2 \times 10^{-16}$ for the null hypothesis.

\section{A Discussion on Group Detection Application}\label{case}

\begin{figure}[!t]
\centering
\subfloat[Nov.2008]{\includegraphics[width=1.5in]{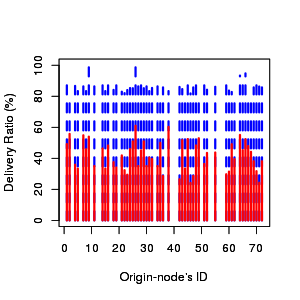}
\label{dr1}}
\hfil
\subfloat[Dec.2008]{\includegraphics[width=1.5in]{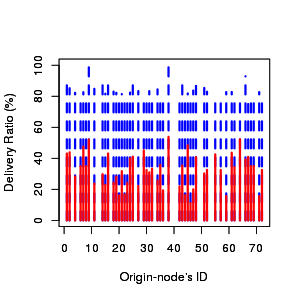}
\label{dr2}}
\hfil
\subfloat[Jan.2009]{\includegraphics[width=1.5in]{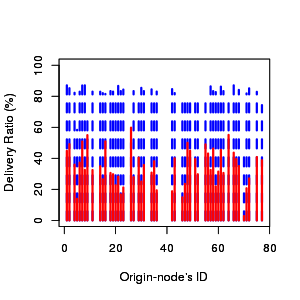}
\label{dr3}}
\hfil
\subfloat[Feb.2009]{\includegraphics[width=1.5in]{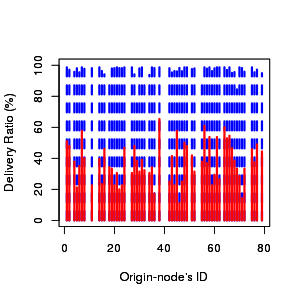}
\label{dr4}}
\hfil
\subfloat[Mar.2009]{\includegraphics[width=1.5in]{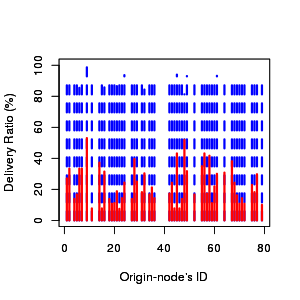}
\label{dr5}}
\hfil
\subfloat[Apr.2009]{\includegraphics[width=1.5in]{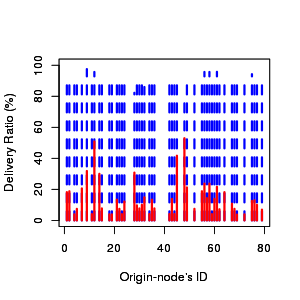}
\label{dr6}}
\hfil
\subfloat[May 2009]{\includegraphics[width=1.5in]{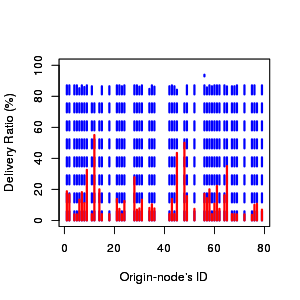}
\label{dr7}}
\hfil
\subfloat[Jun.2009]{\includegraphics[width=1.5in]{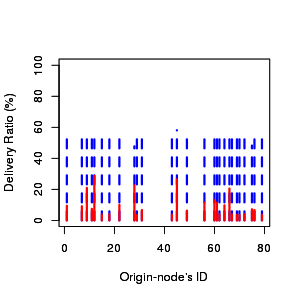}
\label{dr8}}
\hfil

\caption{Delivery ratios for nodes that have been in groups with the origin node (in blue) and for nodes that have not been (in red). Numbers in the x-axis represent IDs of origin nodes}\label{dr}

\end{figure}

\begin{figure}[!t]
\centering
\includegraphics[width=\columnwidth]{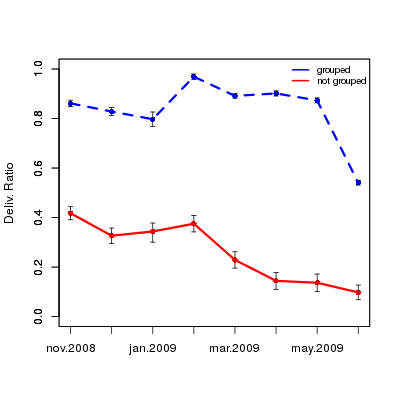}
\caption{Average delivery ratios, for the different origin nodes, from November-2008 to June-2009}
\label{deliv_month}
\end{figure}

There are several applications that can benefit from the knowledge of intrinsic characteristics governing social groups' behavior, for example, DTN protocols. In DTNs, two very important properties are high message delivery ratio and low message delivery times. With the purpose of illustrating how the knowledge about previous group meetings could be used to improve DTN protocols, we here analyze the difference in delivery ratios of nodes that have been in the same group with the source of a given message and of nodes which have not been in the same group of the source.

In our experiment, we select a node as the origin of a message and simulate an epidemic message transmission, i.e., at every time that a node which has the message meets a node that does not have it yet, the message is propagated. We simulate the message propagation selecting each node of the data set as origin and divide the rest of the nodes in two classes: nodes that have belonged to a group together with the origin in the past 30 days and nodes that have not. Then, we compute the delivery ratios of the two classes of nodes. We consider that the message is delivered to node $N$ if node $N$ receives the message within seven days after the start of the dissemination.

As presented in Fig.\ \ref{dr}, for different months, delivery ratios to nodes that had been in group with the origins are more than two times higher. Around 90\% of them receive the message sent by the origin within one week. On the other hand, the delivery ratio for nodes that have not been in a group together with the origin is around 40\%. This result conforms with the periodical behavior discussed in Section \ref{char} (a group that have met in recent past is likely to meet again soon) and is a key insight on how group detection could and should be used to better design opportunistic routing protocols, as discussed in future works. One may notice that, throughout different months, for some node IDs there are some blank spaces in the graphs of Fig.\ \ref{dr}. These are origin nodes that were not active in the data set during that given month, and for this reason present 0\% delivery ratios to both classes of nodes. In June of 2009 for example, there are several nodes with 0\% delivery ratios, which makes sense since many of the students start to leave the campus for summer vacation. Figure \ref{deliv_month} depicts the average delivery ratios, with different origins, from November of 2008 to June of 2009.

\section{Conclusion}\label{conclusion}

In this work, we go over a sequence of methodological steps to detect and track social groups in mobility traces. We perform a characterization of groups' evolution over time considering (i) size; (ii) structure change rates of growth, contraction, birth and death; (iii) group meeting periodicity, and (iv) group meeting durations and its correlation with the strength of group's bonds and group's composition stability.

Our results show that social groups' characteristics are highly dependent on day time. Moreover, group contacts happen in a periodical fashion, presenting not only daily periodicity, but also a portion of weekly periodicity. It is also noteworthy that groups re-encounter probability decreases over time, meaning that groups that haven't met in a while are less likely to meet again soon. Finally, the duration of group meetings is moderately correlated with the stability of it's members ($\rho$) and with the strength of their social bonds ($GSCF$).

As future work, we plan to use unraveled characteristics to design social-groups-aware opportunistic routing and information diffusion protocols that improve delivery ratio and time, decreasing message duplication and network overhead. We also plan to apply the proposed methodology to other proximity traces such as \cite{dartmouth,infocom,usd}.

\ifCLASSOPTIONcaptionsoff
  \newpage
\fi

\bibliographystyle{IEEEtran}
\bibliography{main}

\begin{IEEEbiography}[{\includegraphics[width=1in,height=1.25in,clip,keepaspectratio]{picture}}]{John Doe}
\blindtext
\end{IEEEbiography}

\end{document}